\documentclass[aip,jcp,reprint,showkeys,superscriptaddress,longbibliography,noeprint]{revtex4-1}
\usepackage[utf8]{inputenc}
\usepackage[english]{babel}

\usepackage[fleqn]{amsmath}
\usepackage{amssymb}
\usepackage[separate-uncertainty=true, multi-part-units=single]{siunitx}
\usepackage[inline,shortlabels]{enumitem}

\newcommand{\blue}[1]{{\color{blue} #1}}

\newcommand\TODO[1]{{\color{purple}{\bfseries TODO} #1}}

\renewcommand\blue[1]{} \renewcommand\TODO[1]{} 

\renewcommand{\vec}[1]{\boldsymbol{#1}}

\newcommand\kB{k_\mathrm{B}}

\usepackage{graphicx,grffile,subfigure}
\graphicspath{{./}}
\newlength\figwidth
\usepackage{url,hyperref}
\usepackage[table,usenames,dvipsnames]{xcolor}
\hypersetup{colorlinks=true, linkcolor=BrickRed,
  urlcolor=blue!50!black, citecolor=blue!50!black}

\usepackage[capitalize]{cleveref}
\crefrangelabelformat{equation}{(#3#1#4)$-$(#5#2#6)}

\makeatletter
\def\paragraph{%
  \@startsection
    {paragraph}{4}{\parindent}{\z@}{-1.5em}%
    {\normalfont\normalsize\itshape}%
}%
\makeatother

\begin{document}


\title{Molecular dynamics of open systems: construction of a mean-field particle reservoir}

\newcommand\FUBaffiliation{\affiliation{Freie Universität Berlin, Institute of Mathematics, Arnimallee 6, 14195 Berlin, Germany}}

\author{Luigi Delle Site}
\email{luigi.dellesite@fu-berlin.de}
\FUBaffiliation

\author{Christian Krekeler}
\FUBaffiliation

\author{John Whittaker}
\FUBaffiliation

\author{Animesh Agarwal}
\FUBaffiliation
\affiliation{Theoretical Biology and Biophysics Group, Los Alamos National Laboratory, Los Alamos, New Mexico 87545, USA}

\author{Rupert Klein}
\FUBaffiliation

\author{Felix H{\"o}fling}
\FUBaffiliation
\affiliation{Zuse Institute Berlin, Takustr. 7, 14195 Berlin, Germany}

\begin{abstract}
%
%
The simulation of open molecular systems requires explicit or implicit reservoirs of energy and particles.
Whereas full atomistic resolution is desired in the region of interest, there is some freedom in the implementation of the reservoirs.
Here, we construct a combined, explicit reservoir by interfacing the atomistic region with regions of point-like, non-interacting particles (tracers) embedded in a thermodynamic mean field.
The tracer molecules acquire atomistic resolution upon entering the atomistic region and equilibrate with this environment,
while atomistic molecules become tracers governed by an effective mean-field potential after crossing the atomistic boundary.
The approach is extensively tested on thermodynamic, structural, and dynamic properties of liquid water.
Conceptual and numerical advantages of the procedure as well as new perspectives are highlighted and discussed.
\end{abstract}


\maketitle

%

\blue{\footnotesize\itshape The first paragraph should summarize the reasons for undertaking the work
and the main conclusions which can be drawn.\par}

\paragraph*{Introduction.}


Molecular dynamics (MD) simulations have proven as a powerful means of investigation of molecular systems for half a century now \cite{chandler:annurev2017}.
The steady increase of computing resources has brought larger system sizes and longer time scales into the scope of direct simulations, which may soon be run in parallel to experiments.
Diverse application fields know about situations where the region of interest exchanges particles with an environment, e.g., droplet evaporation \cite{heinen:jcp2016}, condensation in porous hosts \cite{hoeft:jacs2015,peksa:mmm2015}, nanoflows \cite{ganti:prl2017}, and chemical reactions in living cells \cite{schoeneberg:nc2017,engblom:pre2018} and catalysts \cite{roa:zpc2018,matera:acs2014}.
Simulations of such open systems face severe limitations as they often depend
on the efficient insertion of molecules into dense
environments\cite{delgado:pre2003, delgado:jcp2008}.{In Molecular
  Dynamics, the insertion of
  particles into dense simulation boxes has been achieved in various ways: by
  extending the phase space with an additional variable that controls the
  insertion and removal of molecules
  \cite{cagin911,cagin912,ji92,weerasinghe94,lynch97}, by an extended
  Hamiltonian with additional fractional particles
  \cite{palmer94,lo95,shroll99,kuznetova99,eslami2007} or by hybrid techniques
  where the insertion and removal of particles in the MD simulations are
  performed with a Monte Carlo approach \cite{boinepalli2003,
    chempath2003,lsal2005}.
Similar recent efforts in the direction of grand canonical MD and the
construction of a particle reservoir can be found in Refs.\citenum{perego2015,ozcan2017,karmar2018}. 
Here, we present a solution in which, different from the abovementioned
schemes,} molecules are inserted dynamically by converting point particles from a reservoir into molecules, building on ideas of the Adaptive Resolution Simulation (AdResS) technique \cite{dellesite:physrep2017,praprotnik:jcp2005,praprotnik:annurev2008,wang:jctc2012}.
The approach may also be seen as a computational magnifying glass that allows one to model and observe at high resolution only the region of acute interest, thereby focusing computational resources on the essential degrees of freedom \cite{krekeler:jcp2018}.
We evaluate the concept's usefulness exemplarily for the simulation of liquid water, based on a number of thermodynamic, structural, and dynamic properties.
So far, the reservoir of point particles primarily fulfills the role of a thermodynamic bath that ensures physical consistency in the high-resolution region.
As a perspective, the reservoir particles may then act as tracers coupled to hydrodynamic fields in a continuum model, which would enable hybrid multi-scale simulations \cite{delgado:pre2003,delgado:jcp2008,petsev:jcp2015,alekseeva:jcp2016,hu:jcp2018}.
The Adaptive Resolution Simulation (AdResS) technique \cite{praprotnik:jcp2005,praprotnik:annurev2008,wang:jctc2012} has been developed in the past years to fulfill the role of such a magnifying glass in molecular dynamics.
In AdResS, the space is divided into two large regions: a high resolution (atomistic) and a low resolution (coarse-grained) region.
The interface of these regions is characterized by a transition region where interactions, via a space-dependent switching function, smoothly change from atomistic to coarse-grained and vice versa.
The method has been applied with success to several challenging systems, including solvation of large molecules \cite{lambeth:jcp2010,sablic:sm2016,zavadlav:sr2017,fiorentini:jcp2017, agarwal:pccp2017,shadrack:ats2018} and ionic liquids \cite{krekeler:pccp2017,shadrack:jcp2018a,shadrack:jcp2018b} to cite a few. 
Recent developments of the method have shown that changing resolution through a space-dependent function slows computational performance and is overall not convenient for the implementation of the algorithm or its transferability from one code to another.
For this reason, the design of a simpler interface was proposed that is based on an abrupt change of resolution without a switching function and that was extensively tested and shown to be accurate and efficient \cite{krekeler:jcp2018}.
In parallel, Kremer and coworkers, within the framework of AdResS utilizing a smooth switching function, proposed to remove the two-body coarse-grained interactions in the coarse-grained region, reducing it to a gas of non-interacting particles \cite{kreis:epjst2015}.
In the current work, we merge both ideas, creating a system with an abrupt interface between a region of atomistic resolution and a thermodynamic reservoir of non-interacting particles.

Note in this context that AdResS simulations require the action of an additional one-body force at the interface, known as the thermodynamic force. Together with a thermostat, it ensures a thermodynamic equilibrium among the differently-resolved regions
such that thermodynamic properties as well as structural and dynamic correlations of the atomistically resolved subdomain match those of a fully atomistic simulation of the entire system.
Such a force was derived from first principles of statistical mechanics and thermodynamics \cite{poblete:jcp2010,fritsch:prl2012,wang:prx2013}, and, together with the thermostat, will be the key ingredient in the construction of the reservoir of non-interacting particles below.
The combined role of the thermostat and the thermodynamic force is that of
creating a mean field in which the non-interacting particles are embedded, so
that density and temperature of the tracers fluctuate around their equilibrium
values.
This requirement will be shown, through a numerical experiment for the case of liquid water at room conditions, to be the essential condition for a proper, grand canonical-like exchange of particles with the atomistic region, as expected by the principles of statistical mechanics of open systems \cite{agarwal:njp2015,dellesite:pre2016,dellesite:physrep2017}.

\paragraph*{AdResS in a nutshell.}

\begin{figure}
\includegraphics[width=\figwidth]{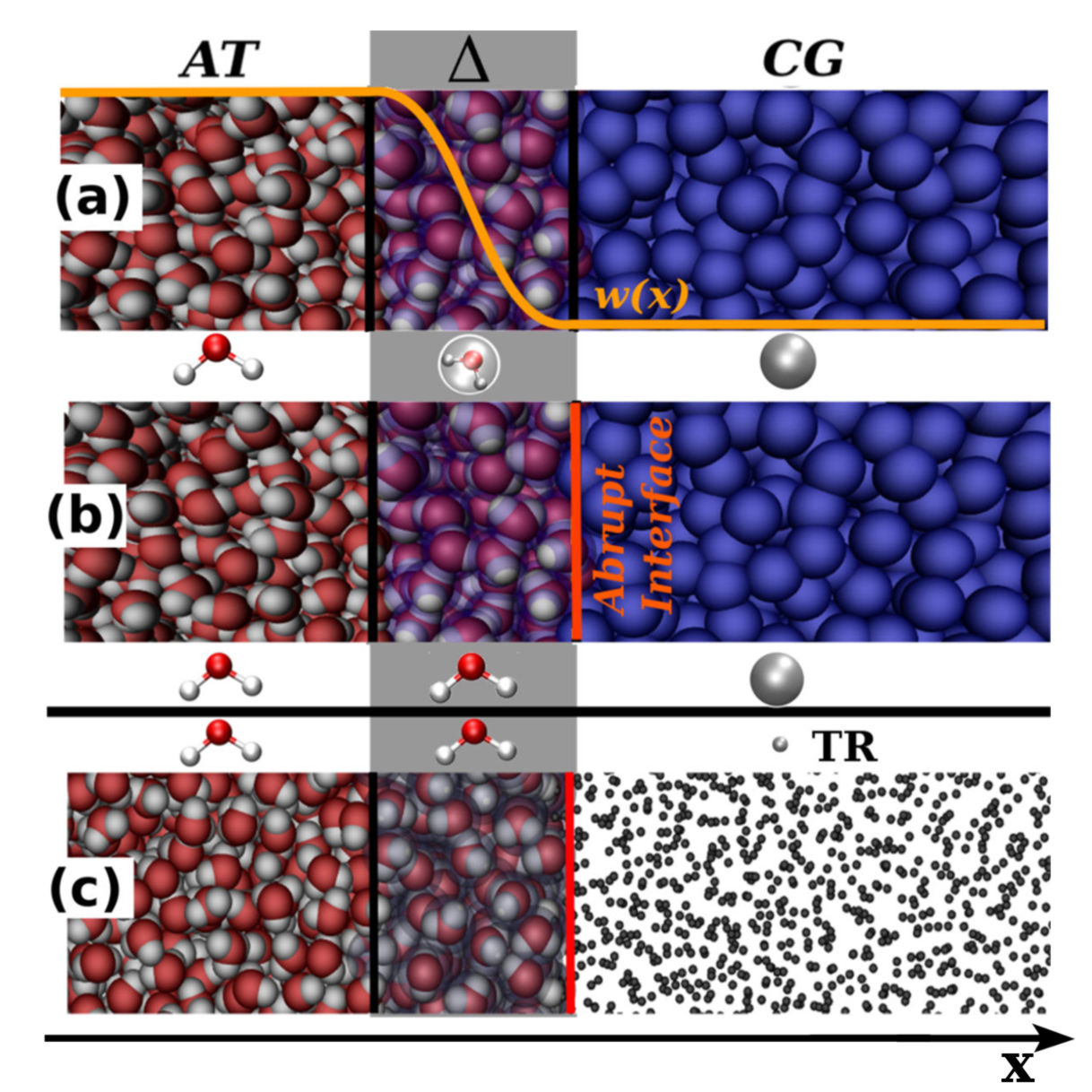}
  \caption{(a) Set-up of the original version of AdResS: in AT,
  molecules have atomistic resolution, in $\Delta$, molecules have hybrid atomistic/coarse-grained
  resolution, according to the switching function $w(x)$, and in CG molecules have coarse-grained resolution.
  ~(b) Set-up of the latest version of AdResS: in the $\Delta$ region a switching function is no more required. Atomistic and coarse-grained resolution molecules are directly interfaced and the $\Delta$ region is large enough to allow molecules to be in equilibrium with the environment before entering into the AT region. Atomistic molecules interact with coarse-grained molecules and vice versa only within a cutoff length equal to the width of the $\Delta$ region along the x-direction via a coarse-grained potential acting on the center of mass of each molecule.
~(c) The AdResS set-up of this work. The atomistic region is directly
  interfaced with a reservoir of non-interacting point-like particles brought
  into equilibrium through the action of the $\Delta$ region. Differently from
  the standard AdResS, where the thermodynamic force is calculated only in
  $\Delta$, in this case the thermodynamic force is calculated over the whole
  $\Delta \cup \mathrm{TR}$ region. {We must note that, for
simplicity, we depict half of our system in the figure. In reality, the AT
region is sandwiched between $\Delta$ and $\mathrm{TR}$ regions on both sides in the $x$-direction. As a consequence only the
    tracers experience the periodic boundary
    condition along the $x$-direction, i.e. the direction along which molecules
  change resolution.}}

  \label{fig:cartoons}
\end{figure}

The original version of AdResS, illustrated in \cref{fig:cartoons}a, defines the coupling between different regions of resolution through a smooth interpolation of the force between pairs of molecules ($\alpha,\beta$):
\begin{equation}
\vec{F}_{\alpha \beta} = w(X_{\alpha})w(X_{\beta})\vec{F}_{\alpha\beta}^\mathrm{AT}
  + [1 - w(X_{\alpha})w(X_{\beta})]\vec{F}_{\alpha\beta}^\mathrm{CG} \,.
\end{equation}
Here, $\vec{F}_{\alpha\beta}^\mathrm{AT}$ is the total force exerted by the atoms of molecule $\beta$ on molecule $\alpha$
due to atomistic interactions;
$\vec{F}_{\alpha\beta}^\mathrm{CG} = -\nabla U^\mathrm{CG}(\vec R_\alpha - \vec R_\beta)$  is the coarse-grained force between the centers of mass of both molecules, $\vec R_\alpha$ and $\vec R_\beta$;
and the switching function $w(X)$ interpolates smoothly between 1 and 0 in the transition region~$\Delta$
with $X=\vec n\cdot \vec R$ denoting the projection of the position $\vec R$ on the interface normal~$\vec n$.
The global thermodynamic equilibrium across the entire simulation box is assured by the thermostat and by the thermodynamic force $\vec F_\mathrm{th}(x)$.
{The latter is a position-dependent force, $\vec F_\mathrm{th}(x) = (M/\rho_0) \nabla p(x)$, that counteracts the pressure gradient between regions of different resolution \cite{fritsch:prl2012}, which would result from entropy differences due to the mismatch in the number of degrees of freedom.}
It is calculated iteratively during the equilibration run from the gradient of the mass density $\nabla \rho^{(k)}(x)$ in the $\Delta$ region at the $k$-th iteration step:
$\vec{F}_{th}^{(k+1)}(x)=\vec{F}_{th}^{(k)}(x) -
c\bigl(M/\rho_0^2\kappa_T\bigr) \nabla\rho^{(k)}(x)$, where $M$ is the mass of
a molecule, $\kappa_T$ the isothermal compressibility, $c$ a constant to
control the convergence rate, and $\rho_0$ the equilibrium density
\cite{poblete:jcp2010,fritsch:prl2012,wang:prx2013,dellesite:physrep2017}. {The
  constant $c$ is tuned by starting from a small value (about 0.0015, i.e. slow convergence)
  and after a certain amount of iterations, stepwise increased. When
  variations become larger than the converge rate of the previous 
step  then it is
  stepwise reduced until final convergence is reached with the criterion: $\frac{max|\rho^{(k)}(x)-\rho_0|}{\rho_0}\le 0.02$.}   
After $\vec{F}_{th}(x)$ has been determined, it remains fixed during the production runs.

In the latest variant of the method, proposed in Ref.~\citenum{krekeler:jcp2018}, the simulation box is divided in only two regions of different resolution, atomistic ($\mathrm{AT} \cup \Delta$) and coarse-grained (CG), with the transition region $\Delta$ using the unchanged atomistic interactions throughout.
The coupling is achieved in a direct way (\cref{fig:cartoons}b) by the pair potential
$U_{\Delta}=\sum_{\alpha\in\Delta}\sum_{\beta\in \text{CG}}U^{CG}(\vec{R}_{\alpha} - \vec{R}_{\beta})$
acting only between pairs of particles at different sides of the $\Delta$/CG
interface.
{The minimum size of the $\Delta$ region is fixed by the cutoff of the intermolecular (non-bonded) interaction.
  This implies that molecules of the AT region do not have any direct interaction with
  the molecules of the CG region and as a consequence the corresponding dynamics is determined only by atomistic
  interactions. The same set up can be used for large molecules, e.g. long
polymers, in which case only the non-bonded interactions are subject to
the adaptive resolution process according to the position of each polymer
segment in the box, while intramolecular interactions remain atomistic in
nature regardless of the segment's position, see e.g. Refs.\citenum{peters2016,ciccotti2019}}.
{Further, the thermodynamic force and the thermostat act only in the $\Delta$ and CG regions,} i.e., the dynamics in the AT region follows the untweaked atomistic Hamiltonian.
In $\Delta$, the thermodynamic force is applied together with a capping force to ensure that the sudden change in a molecule's resolution does not generate unphysically large forces before a local equilibration is established (i.e., avoiding that a CG molecule crosses the CG/AT interface, instantaneously gains atomistic features, and is found in an undue overlap configuration with surrounding atomistic molecules).
The evaluation of observables is restricted to the $\text{AT}$ region, i.e., molecules in the $\Delta$ region are excluded as this region is contaminated by artifacts of the coupling.
This simplified approach was tested with success in the case of both liquid water and ionic liquids \cite{krekeler:jcp2018}.
The natural question that arises is whether the basic characteristics of molecular adaptivity, used in Ref.~\citenum{krekeler:jcp2018}, can be further reduced, as suggested by the preliminary work of \citet{kreis:epjst2015}.
In the following, we describe the merging of the ideas of Refs.~\citenum{krekeler:jcp2018} and \citenum{kreis:epjst2015} and discuss the horizons that such an approach opens.

\paragraph*{From interacting coarse-grained particles to non-interacting tracers.}

\Cref{fig:cartoons}c shows the set up for the AdResS simulation with the abrupt interface, a transition region $\Delta$ with atomistic resolution, and the coarse-grained region containing non-interacting particles that we refer to as \emph{tracers} (TR).
Thermodynamic equilibrium is assured by a thermostat, acting on all particles in the $\Delta$ and TR regions, and by the thermodynamic force $\vec F_{th}(x)$.
This latter quantity is calculated in the iterative manner as explained above, however, now we lift its restriction to the $\Delta$ region and let it extend into the TR region. In fact we do not need anymore the condition that in the CG region the potential acting on the particles should be, for consistency, the coarse-grained potential only.
Let us introduce the corresponding one-body potential:
$\phi(x)=-\int_{x_0}^x \vec{F}_{th}(x') \cdot \vec{e}_x \,\mathrm{d}x'$ with $x_0$ the position of the AT/$\Delta$ interface.
The total potential energy of the system is
\begin{equation}
  U_\mathrm{tot}= U_\mathrm{tot}^\mathrm{AT}
    + \sum_{k\in \Delta \cup \mathrm{TR}}\phi(X_k) \,,
\end{equation}
where $U_\mathrm{tot}^\mathrm{AT}$ is the total potential energy due to atomistic force fields in the AT region
and the $k$-sum runs over all molecules in $\Delta$ and all tracers.
To avoid numerical instabilities, a capping force is employed in $\Delta$ for molecules which, upon just having entered the atomistic region, experience forces that are orders of magnitude less probable than what is typical in the AT region;equivalent measures are found in Refs.~\citenum{krekeler:jcp2018} and \citenum{kreis:epjst2015}.
The capping is achieved by simply truncating each Cartesian component of the \emph{total} force vector at a prescribed maximum value.
The capping occurs only upon insertion of atomistic molecules, i.e., in the vicinity of the border to the TR region, and, despite being a technical artifact, its repercussions on the AT region are not noticeable.
In summary, the action of the environment on each tracer depends only on the specific position of the tracer in space, independently from the overall tracers configuration, that is in $\mathrm{TR} \cup \Delta$ region each tracer experiences $\phi(x)$ and the thermostat as an effective \emph{mean field}.

\paragraph*{Test on Liquid Water.}

We have tested the proposed scheme in simulations of liquid water at room conditions; technical details can be found at the end.
The reference setup consisted of \num{2.7e4} SPC/E water molecules at ambient conditions within an elongated cuboid box of length $L_x = \SI{60}{nm}$ and cross-section $\approx (\SI{3.7}{nm})^2$.
Van der Waals and electrostatic interactions were cut off at a distance of $r_c = \SI{1.2}{nm}$.
In the corresponding AdResS setup, the atomistic resolution is kept only in a small region of length $L_\mathrm{AT} = \SI{3}{nm}$ (along the $x$-axis) expanded by transition regions $\Delta$ at both ends of length $L_\Delta = \SI{1}{nm}$.
The remaining volume TR ($L_\mathrm{TR} = \SI{55}{nm}$) is filled with an ideal gas of tracer particles, with each tracer representing one water molecule;
the mass density of tracers is fixed to its atomistic value, here $\rho_0 = \SI{e3}{kg.m^{-3}}$.
Employing the ideal gas law, we obtain a pressure in the TR region of
$p_\mathrm{TR} = (\rho_0/M) \kB T \approx \SI{1,400}{bar}$.
Thus as with many coarse-grained models \cite{fritsch:prl2012,kreis:epjst2015,alekseeva:jcp2016,roy:jcp2016}, the pressure is orders of magnitude larger than in the corresponding atomistic model, here $p_\mathrm{AT} = \SI{1}{bar}$.
The corresponding isothermal compressibility $\kappa_T^\mathrm{(TR)} = 1/p_\mathrm{TR}$ exceeds that of water in the AT region by merely a factor of 16.
We have intentionally chosen a large reservoir of tracers and a small transition region, $L_\Delta \ll L_\mathrm{TR}$.
The idea behind this choice is to test the computational stability of the algorithm.
In fact, in a setup with a smaller TR region and larger AT and $\Delta$ regions, the extended domain with atomistic interactions is likely to more easily stabilize the system towards the wished target, which is fully atomistic.
Instead, if for a large TR region and relatively small $\Delta$ and AT region, $\phi(x)$ and the thermostat can equilibrate such a large system and lead to accurate results, then one can be confident that the method is computationally stable.
In fact, in the setup employed in Ref.~\citenum{kreis:epjst2015}, where a relatively small tracer region was coupled to a large atomistic region ($\mathrm{AT} \cup \Delta$) the system was stabilized essentially by the fully atomistic part.
In order to remove any doubt about possible artifacts arising from the elongated (tube-like) shape of the simulation box, we have tested also a setup where the AT region is defined by a small sphere embedded in a large cube filled with tracers.
The obtained structural properties display the same accuracy of the original tube-like setup.

\begin{figure}
  \centering
  \includegraphics[width=\figwidth]{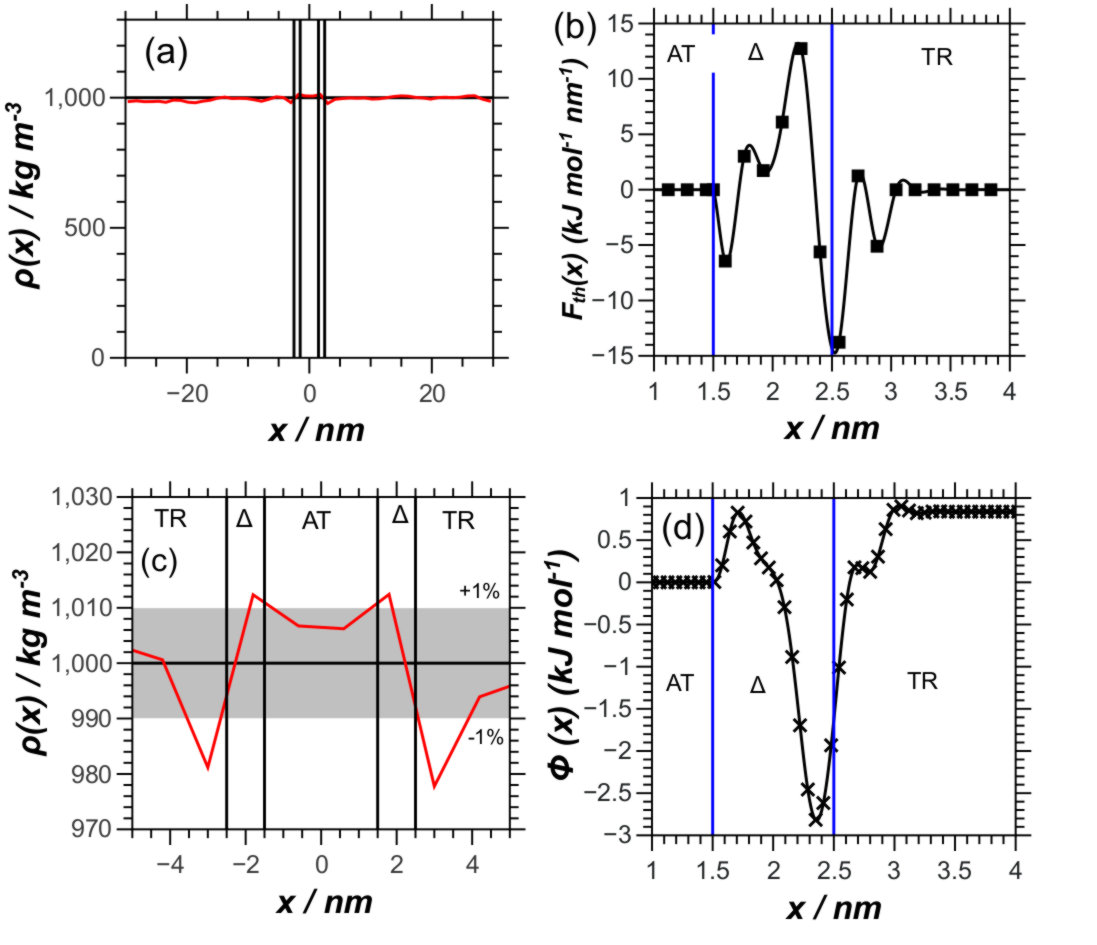}
  \caption{Panels~a,c: Mean profiles of the particle density across the entire box, panel~c shows a zoomed view of the central region. The gray band indicates the accuracy goal of 1\% deviation from the target value.
  ~Panel~b: Thermodynamic force $\vec F_{th}(x)$ along the $x$-axis in the region of major interest. By construction, $\vec F_{th}(x)=0$ in the AT region. Symbols are actual data, and the line shows a smooth interpolation.
  ~Panel~d: Thermodynamic potential $\phi(x)$ corresponding to the force data of panel~b.}
  \label{fig:profiles}
\end{figure}

In the following, we evaluate the new coupling scheme based on the mean density profile and various pair and auto-correlation functions.
The mean density profile $\rho(x)$ obtained from AdResS production runs (\cref{fig:profiles}a) is flat across the entire simulation box and reproduces the equilibrium value $\rho_0$ of the atomistic reference simulation within 1\% (\cref{fig:profiles}c), except very close to the $\Delta$/TR interface, where the relative deviation is 2\%.
These accuracy goals were chosen by us in the equilibration run as convergence criterion for the determination of the thermodynamic force.

\begin{figure*}
  \centering
  \includegraphics[width=\textwidth]{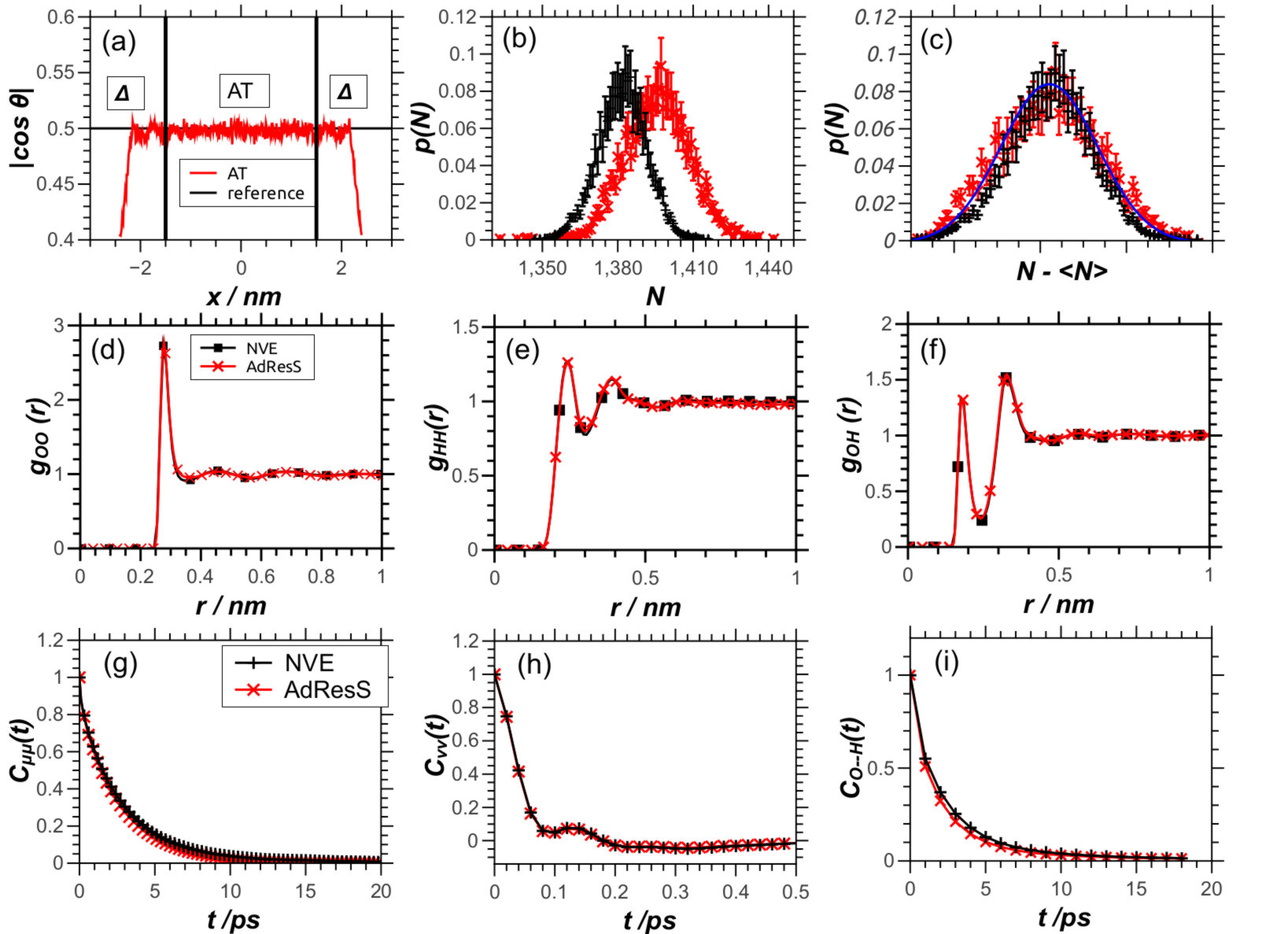}
  \caption{
    Comparison of thermodynamic, structural, and dynamic properties of the AT region of the present AdResS approach (red data points) with a fully atomistic reference simulation (black).
    ~Panel a: mean profile of the orientation of the water dipole in the AT and $\Delta$ regions; the angle $\theta$ is measured relative to the $x$-axis.
    ~Panel b: probability distribution $p(N)$ of the particle number in the AT region (red) compared with results from the equivalent subdomain of the fully atomistic reference (black).
    ~Panel c: same data as in panel~b, corrected for the slight discrepancy in the peak values. The blue line indicates the fit to a Gaussian distribution.
    Panels d--f: radial distribution functions for oxygen--oxygen (d),
    hydrogen--hydrogen (e), and oxygen--hydrogen (f) atoms. Curves are smooth interpolations between the data points.
    ~Panels g--i: dynamic autocorrelation function of the electric dipoles (g) and the center-of-mass velocity (h) of a water molecule; the hydrogen bond correlation is shown in (i). Correlation functions have been normalised to $C(0)=1$.
  }
  \label{fig:properties}
\end{figure*}

The thermodynamic force $\vec F_{th}(x)$, used throughout in the production run, is plotted in \cref{fig:profiles}b with the corresponding thermodynamic potential $\phi(x)$ shown in \cref{fig:profiles}d.
The latter exhibits a barrier next to the AT region and, towards the TR region, a larger trap of depth $1.1\,\kB T$.
In addition, we obtained some fine structure in $\phi(x)$, which is needed to achieve the small roughness of the density profile $\rho(x)$.
In the AT region, $\vec F_{th}(x)=0$ and $\phi(x)=0$ by construction, and $\vec F_{th}(x)$ vanishes rapidly in the TR region, within less than the cutoff length $r_c$.
Hence, the potential $\phi(x)$ rapidly approaches a constant $\phi_\mathrm{TR}$ as the distance to the $\Delta$ region increases.
Tracer particles that enter the atomistic region gain an energy difference of
{about $\Delta\phi := \phi(x_{\Gamma_{TR}})- \phi(x_{\Gamma_{AT}})$, here
  $x_{\Gamma_{TR}}$ means that the $x$-position of the particle is at the interface, $\Gamma_{TR}$, between the TR region
  and the $\Delta$ region, and, equivalently, $x_{\Gamma_{AT}}$ means that the $x$-position of the particle is at the interface, $\Gamma_{AT}$, of the AT region with the $\Delta$ region. Interestingly,} $\Delta\phi$ compensates
exactly the pressure difference between the AT and TR region,
$\Delta \phi = (M/\rho_0) (p_\mathrm{TR} - p_ \mathrm{AT})$.
Due to the high pressure of the TR region and making use of the ideal gas law, this reduces to
$\Delta \phi \approx (M/\rho_0) p_\mathrm{TR} = \kB T$,  that is $\Delta\phi \approx \SI{2.5}{kJ/mol}$ at $T=\SI{300}{K}$.
A measure for the orientation of water molecules is given by their electric dipole. The spatial profile of the dipole orientation matches that of a uniform liquid in the whole AT region, which is highly satisfactory (\cref{fig:properties}a); merely boundary effects are present towards the $\Delta/\mathrm{TR}$ interface due to the absence of molecular dipoles in the TR region, yet this is outside of the atomistic system of interest.
We conclude the action of the thermodynamic force is concentrated in the transition region $\Delta$, while the thermostat in the TR and $\Delta$ regions is sufficient to keep the rest of the large tracer region in equilibrium.

We checked further that the passage of molecules through the transition region is not hindered and that there is a proper exchange between the AT and TR regions. Qualitatively, this is confirmed by coloring all molecules according to their initial region and monitoring their diffusion as time progresses, see supporting figures \labelcref{fig:diffusion_profiles,fig:diffusion_cartoon}.
As a more stringent test, we analyzed the fluctuations of the particle number in the AT region
and compared their probability distribution $p(N)$ with that of the equivalent subregion in a fully atomistic reference simulation, see \cref{fig:properties}b.
The peak values, $N_\mathrm{AT} \approx 1{,}400$, differ slightly by about 1.5\% due to marginal differences in the density (\cref{fig:profiles}c).
Yet, after correcting for this, both distributions neatly collapse and approximately resemble a Gaussian distribution (\cref{fig:properties}c).

Detailed structural properties in the atomistic region have been verified in terms of the radial distribution functions
$g_\mathrm{OO}(r)$, $g_\mathrm{OH}(r)$, and $g_\mathrm{HH}(r)$ of oxygen and hydrogen atoms.
The results from the adaptive simulation are virtually indistinguishable from the fully atomistic reference, see \cref{fig:properties}d--f; in particular, all maxima and minima of the curves are very well reproduced.
Eventually, we have tested the dynamics in the AT region by computing the molecular dipole-dipole autocorrelation functions, $C_{\mu\mu}(t)$, and the velocity-velocity autocorrelation function, $C_{vv}(t)$,  as well as the hydrogen bond correlation function $C_{\mathrm{O}-\mathrm{H}}(t)$ (\cref{fig:properties}g--i) (for the specific definitions of the aforementioned correlation functions see \cite{agarwal:njp2015}.
The agreement with the results from the fully atomistic simulation is rather satisfactory.


\paragraph*{Conclusions.}

We have constructed a hybrid simulation scheme that couples a small open system of atomistic resolution with a thermodynamic bath of point particles (tracers) such that both subsystems are in thermodynamic equilibrium. The reservoir is characterized by its temperature and mass density, which can be viewed as a thermodynamic mean field the tracers are embedded into while displaying their canonical fluctuations.
In the atomistic region, we have shown on the example of liquid water that the scheme reproduces thermodynamic, structural, and dynamic properties with high fidelity.
Higher accuracy can be achieved with finer parametrizations of the thermodynamic force and a tighter convergence criterion for the resulting density profile.
In practice, the huge TR region used here by purpose can be replaced by a much smaller volume as there are no spatial correlations between the tracers.

The new method provides a stark simplification over previous AdResS setups:
\begin{enumerate*}[(i),font=\itshape, itemjoin=\quad]
  \item it avoids the smooth force interpolation in the transition region,
  \item there is no need for double resolution of molecules due to the absence of a coarse-grained potential,
  \item it uses a computationally trivial reservoir.
\end{enumerate*}
Thus, our approach opens a perspective towards efficient non-equilibrium simulations, which would benefit from the easy insertion of reservoir particles. Further, the simple structure of the coupling suggests extensions that satisfy additional constraints, such as momentum conservation, with the ultimate goal of coupling the atomistic region to fluid-mechanical continuum fields.

\section*{Simulation details}

The atomistic reference system consists of 27,648 SPC/E water molecules in a cuboid box of dimensions $\SI{60.00}{nm} \times (\SI{3.72}{nm})^2$ at temperature $T = \SI{300}{K}$, mass density $\rho_0 = \SI{e3}{kg.m^{-3}}$, and pressure \SI{1}{bar}.
Simulations were performed with GROMACS \cite{gromacs} 5.1.0 in the NVE ensemble with a time step of \SI{2}{fs}.
The van der Waals and electrostatic interactions were treated with the ``switch'' method and the reaction-field method, respectively \cite{agarwal:njp2015}, and the cutoff radius for each was set to $r_c = \SI{1.2}{nm}$ as in our previous study \cite{krekeler:jcp2018}.
After equilibration, observables were recorded from one production run over \SI{10}{ns}.

The corresponding AdResS simulations were run with a modified and extended GROMACS code, with the following essential differences to the reference setup:
the elongated box is divided along the long $x$-axis in one AT region ($L_\mathrm{AT} = \SI{3.0}{nm}$), two $\Delta$ regions with $L_\Delta = \SI{1}{nm}$, and two large TR regions connected through periodic boundaries ($L_\mathrm{TR} = \SI{55}{nm}$ in total). Thus, on average only 2,300 water molecules are treated atomistically ($\mathrm{AT} \cup \Delta$ region).
For tracers entering the $\Delta$ region, atomistic forces are capped at a threshold of \SI{2,000}{kJ.mol^{-1}.nm^{-1}}.
The stochastic dynamics integrator (Langevin thermostat) was applied in the
$\Delta \cup \mathrm{TR}$ region with a relaxation time constant of
\SI{0.05}{ps} and a time step of \SI{2}{fs}, while in the AT region the
thermostat part was switched off. {Here we have employed the simplest
  option of a single thermostat acting on $\Delta \cup \mathrm{TR}$, however
  the option of thermostating the two regions separately may be technically
  more efficient (see e.g. Ref.\citenum{praprotnik-thermo2008}) and  it will be
  explored in future work.} 
The thermodynamic force $\vec F_{th}(x)$ points along the $x$-axis by symmetry and calculated from cubic splines parametrized on a uniform grid of spacing \SI{0.16}{nm}. It is obtained iteratively during the equilibration procedure, which is interrupted every \SI{1}{ns} to readjust $\vec F_{th}(x)$; after 30 iterations, the desired flatness of the density profile $\rho(x)$ of 1\% was achieved.
The AdResS production run was then performed with the converged thermodynamic force for a duration of \SI{10}{ns}.


\acknowledgments{We are grateful to Matej Praprotnik for useful discussions. This research has been funded by Deutsche Forschungsgemeinschaft (DFG) through the grant CRC 1114: ``Scaling Cascades in Complex Systems'', project C01. This work has also received funding from the European Union's Horizon 2020 research and innovation program under the grant agreement No. 676531 (project E-CAM).The simulations were performed with the HPC resources provided by the
North-German Supercomputing Alliance (HLRN), grant no. bec00171}

\bibliography{tracers}

\appendix
\clearpage

\section*{Supporting material}

\renewcommand{\theequation}{S\arabic{equation}}
\renewcommand{\thefigure}{S\arabic{figure}}
\renewcommand{\thetable}{S\arabic{table}}
\renewcommand{\theHequation}{S\arabic{equation}}
\renewcommand{\theHfigure}{S\arabic{figure}}
\renewcommand{\theHtable}{S\arabic{table}}
\setcounter{equation}{0}
\setcounter{figure}{0}
\setcounter{table}{0}

\begin{figure}[b]
  \centering
  \includegraphics[width=\figwidth]{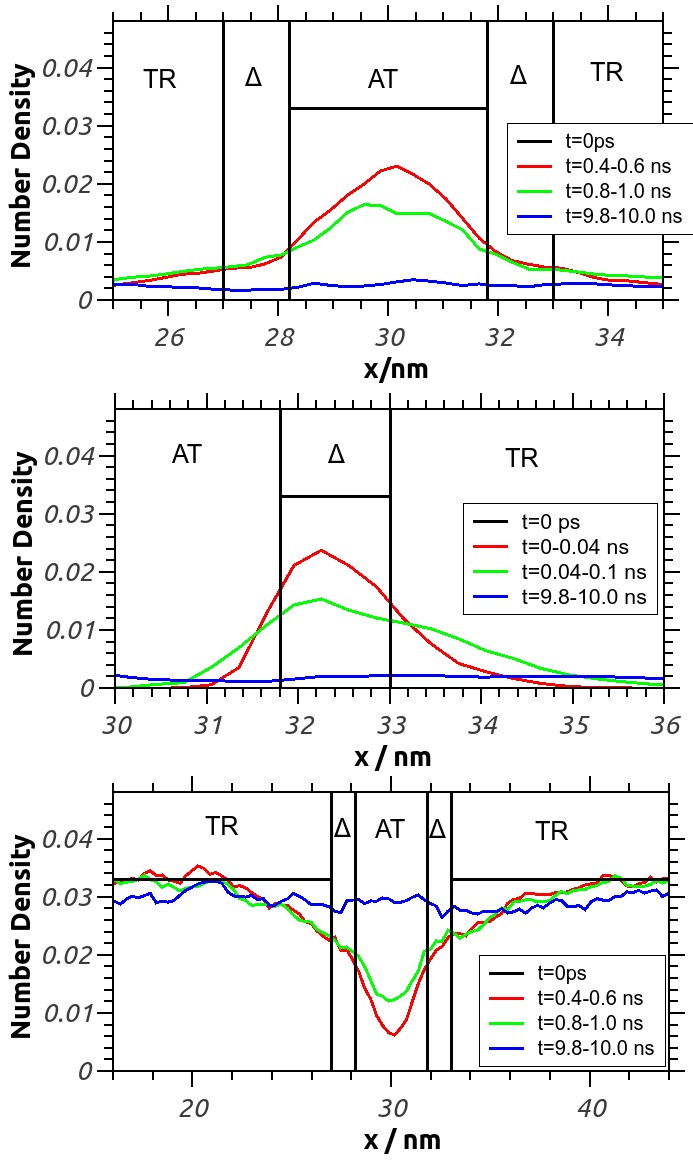}
  \caption{Diffusion of a sample of particles taken at $t=0$ in
    each region and followed in their passage through the other
    regions as time progresses. The diffusion profile is as expected.}
\label{fig:diffusion_profiles}
\end{figure}

\begin{figure}
  \centering
  \includegraphics[clip=true,trim=0cm 0cm 0cm
  0cm,width=9cm]{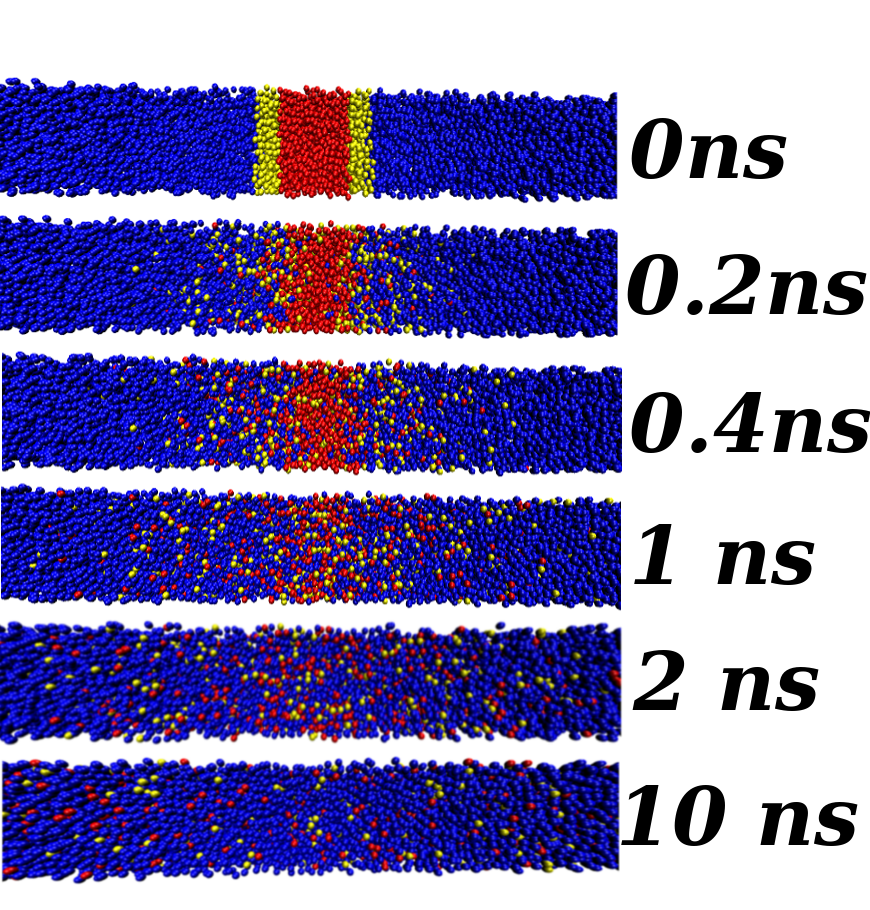}
  \caption{Visualization of the diffusion of particles across the whole simulation box from snapshots of the trajectory at different time steps. Red particles are initially located in the AT region, yellow particles in $\Delta$, and blue particles are in the TR region. After 10 ns the distribution becomes uniform, as anticipated from \cref{fig:diffusion_profiles}.}
  \label{fig:diffusion_cartoon}
\end{figure}

\end{document}